\documentclass[a4paper,10pt]{article}
\usepackage[utf8]{inputenc}
\usepackage[russian]{babel}
\usepackage{amsmath}
\usepackage{amsfonts}
\usepackage{amssymb}

\title{Авторезонансный солитон и накачка с убывающей амплитудой}
\author{О.М. Киселев}

\begin{document}
\maketitle
\begin{abstract} 
Выведено уравнение главного резонанса в частных производных в одномерной среде при воздействии внешней силы с медленно меняющейся частотой -- нестационарное нелинейное уравнение Шредингера с внешней силой. При $t\to\infty$ построен главный член локализованного в пространстве растущего асимптотического решения с солитонным профилем в главном порядке. Оказалось, что растущее по времени асимптотическое  решение может быть получено для внешнего возбуждения с убывающей амплитудой. Выведены необходимые условия роста такого решения при диссипации.
\end{abstract}

\par
В работе исследуется асимптотическое решение уравнения 
\begin{equation}
-i\partial_{\tau}\Psi+\partial^2_{\zeta}\Psi+\left(|\Psi|^2-\tau\right)\Psi + F-i\frac{\nu}{2}\Psi=0. \label{primResEq}
\end{equation}
Это уравнение определяет огибающую одиночной моды плоской волны малой амплитуды в слабо нелинейной среде с внешним осциллирующим возбуждением, частота которого линейно меняется. Это уравнение можно рассматривать как обобщение нелинейного уравнения Шредингера \cite{Kelley,Talanov,Zakharov}, а также как аналог  известного обыкновенного уравнения главного резонанса \cite{FajansFriedland,KalyakinReview} и как нелинейное обобщение уравнения локального резонанса \cite{Neu}.
\par
С точки зрения приложений в физике важно изучить асимптотические свойства решений этого уравнения при больших $\tau$ и выделить растущие по $\tau$ и локализованные по $\zeta$ решения, если таковые существуют. В ситуации общего положения существование растущих решений открывает путь к генерации уединенной волны конечной амплитуды на авторезонансе. Ранее были получены асимптотические формулы для генерации огибающей в виде солитона нелинейного уравнения Шредингера на локальном резонансе \cite{GKL,GKT}. Для авторезонанса  известны в большей степени численные результаты по генерации кноидальных волн близких к солитонам \cite{FriedlandShagalov,Friedland-Shagalov} и по управлению параметрами солитона как конечной, так и малой амплитуды с помощью внешней накачки \cite{Maslov-Kalyakin-Shagalov,KalyakinGarifullinShamsutdinov,Garifullin}.
\par
Схему вывода уравнения (\ref{primResEq}) приведем, воспользовавшись уравнением sine-Gordon. Уравнение sine-Gordon является популярной моделью при исследовании уединенных волн  в различных нелинейных средах с дисперсией. Учет внешних возмущений среды приводит к различным вариантам возмущений уравнения sine-Gordon. В предлагаемой работе изучается зарождение уединенной волны -- бризера в возмущенном уравнении:
\begin{equation}
u_{tt}-u_{xx}+\sin(u)=\varepsilon^3 f\cos(S)-\varepsilon^3\mu u_t,\quad 0<\varepsilon\ll1.\label{ppSG}
\end{equation}
фаза возмущения  $S=kx+\omega t-\varepsilon^2 t^2/2$, где $\omega^2=k^2+1$, $k$ -- произвольный параметр. Амплитуда возмущения -- $f$.
\par
Возмущенное уравнение (\ref{ppSG}) учитывает, например, слабое внешнее воздействие. В присутствии солитонов в ситуации общего положения параметрическое возмущение приводит к изменению  параметров уединенных волн на больших временах. 
\par
Результат работы состоит в следующем -- построен главный член растущего по $\tau$  и локализованного по $\zeta$ асимптотического решения уравнения (\ref{primResEq}) при убывающей как $1/\sqrt{\tau}$ амплитуде накачки $F$. Аналог этого асимптотического решения  для обыкновенного уравнения главного резонанса 
\begin{equation}
i\psi'+(|\psi|^2-\tau)\psi=F,
\label{oPrimResEq}
\end{equation}
соответствует специальному решению:
\begin{equation}
\psi=\sqrt{\tau}e^{ia},\quad F=ie^{ia}/(2\sqrt{\tau}),\quad a\in \mathbf{R}.
\label{oPrimResSol}
\end{equation}

\section{Мотивировка}
\par
Для линейного обыкновенного уравнения
$$
y''+y=g\cos(t+b)
$$
скорость роста резонансных решений определяется интегралом от амплитуды накачки $g(t)$. 
\par
Известно, что существуют авторезонансные решения уравнения  (\ref{oPrimResEq}) растущие как $\sqrt{\tau}$ при $F=const$, см, например, обзор \cite{KalyakinReview}.
\par
Удивительный факт -- существование решения (\ref{oPrimResSol} уравнения главного резонанса (\ref{oPrimResEq}), которое растет так же быстро, как авторезонансные решения уравнения с постоянной, неубывающей накачкой $F$. По сути, решения уравнения (\ref{oPrimResEq}) при $F=const$ можно рассматривать как возмущения специального решения (\ref{oPrimResSol}).
\par
Цель предлагаемой работы -- построить солитонный аналог специального растущего решения для (\ref{primResEq}), который бы являлся главным членом авторезонансной асимптотики.

\section{Вывод модельного уравнения для авторезонанса}
\par
В этом разделе приведена схема вывода нелинейного уравнения солитонного авторезонанса. Эта схема повторяет известный формальный вывод нелинейного уравнения Шредингера для слабонелинейных волн \cite{Kelley,Talanov,Zakharov}. Приведенные ниже выкладки объясняют появление неавтономного слагаемого в нелинейном уравнении Шредингера и связи этого слагаемого с модулированной частотой внешнего возмущения уравнения (\ref{ppSG}). Математическое обоснование приведенной здесь схемы возможно при педантичном следовании \cite{KalyakinReview1}. 
\par
Асимптотическое решение с малой амплитудой будем строить с помощью метода многих масштабов. Будем считать переменные $x,t$ быстрыми, а переменные $t_1=\varepsilon t$, $x_1=\varepsilon x$, $t_2=\varepsilon^2 t$ -- медленными. Для этого введем быстрые и медленные переменные
\begin{equation}
u(x,t,\varepsilon)\sim\varepsilon u_1(x,t,x_1,t_1,t_2,\varepsilon) +\varepsilon^2 u_2(x,t,x_1,t_1,t_2,\varepsilon)+\varepsilon^3 u_3(x,t,x_1,t_1,t_2,\varepsilon).\label{as}
\end{equation}
Зависимость от быстрых переменных  и медленных  удобно разделить, оставив зависимость от быстрых переменных и малого параметра только в в фазовой переменной $S$: 
$$
u_1(x,t,x_1,t_1,t_2,\varepsilon)=A(t_1,x_1,t_2)\exp(iS)+c.c.,
$$
где $c.c.$-- комплексно сопряженное слагаемое.
\par
При $\varepsilon^2$ получается неоднородное уравнение для $u_2$:
\begin{equation}
\partial_{tt}u_2-\partial_{xx}u_2+u_2=(-i\omega \partial_{t_1}A +ik \partial_{x_1}A)\exp(iS)+ c.c..\label{u2Eq}
\end{equation}
Зависимость от быстрых переменных в множителе $\exp(iS)$ из правой части этого уравнения содержит решение однородного уравнения для $u_2$. Такая правая часть приводит к растущим по быстрой переменной решениям уравнения для $u_2$. Требование ограниченности $u_2$ по быстрым переменным приводит к уравнению для $A$:
\begin{equation}
\omega \partial_{t_1}A -k \partial_{x_1}A =0.\label{psiEq1}
\end{equation}
Решение этого уравнения -- произвольная функция от характерной переменной $\xi_1=kt_1+\omega x_1$. В результате уточнена зависимость $u_1$ от медленных переменных $x_1$ и $t_1$: $A=A(\xi_1,t_2)$.
\par 
В порядке $\varepsilon^3$ получается уравнение:
\begin{equation}
\partial^2_{tt}u_3-\partial^2_{xx}u_3+u_3=f\cos(S) + \frac{1}{6}u_1^3  -2i\omega\partial_{t_2}u_1 +\partial^2_{\xi_1} u_1 - 2t_2\omega u_1-i\nu\omega u_1.\label{u3Eq}
\end{equation}
Для ограниченности $u_3$ по быстрой переменной $t$ необходимо исключить слагаемые, соответствующие решениям однородного уравнения из правой части. В результате получается уравнение для функции $A$.
\begin{equation}
-2i\omega\partial_{t_2}A+\partial^2_{\xi_1}A+\left(-2\omega t_2 +{1\over2}|A|^2\right)A + \frac{1}{2}f-i\nu\omega A=0. \label{primResEq2v}
\end{equation}
Переобозначим $t_2=\tau$ и сделаем замены: $\sqrt{2\omega}\xi_1=\zeta$, $A(t_2,\xi_1)=2\sqrt{\omega}\Psi(\tau,\zeta)$, $F=f/(8\sqrt{\omega^3})$. В результате получится уравнение (\ref{primResEq}). Это уравнение является обобщением нелинейного уравнения Шредингера для авторезонанса с диссипацией.

\section{Асимптотическое решение модельного уравнения}
\par
Для того, чтобы убедиться, что это уравнение действительно является модельным для солитонного авторезонанса, необходимо показать, что оно допускает растущие решения с солитонным профилем.
\par
Для исследования растущих при $\tau\to\infty$ решений удобно сделать замены:
$$
\overline{\Psi}=\sqrt{2\sqrt{2\sigma}}\phi(\sigma,z)\exp(-i\sigma),\quad\sigma=\tau^2/2,\quad z=\sqrt[4]{2\sigma}\zeta.
$$
Подставим эту формулу в уравнение. После несложных преобразований получим:
\begin{eqnarray}
i\partial_\sigma\phi+\partial_{zz}\phi+2|\phi|^2\phi+\frac{\cal F}{\sigma^{3/4}}\exp(i\sigma) +{i\over4\sigma}\phi=0,
\label{perturbedNShE}
\end{eqnarray}
где ${\cal F }=\overline{F}/(2\sqrt[4]{2})$.
\par
В результате построение авторезонансного решения при $\sigma\to\infty$ сводится к исследованию решений возмущенного нелинейного уравнения Шредингера. Построим неубывающее асимптотическое решение при $\sigma\to\infty$. Для простоты будем строить решение с односолитонным главным членом. При построении удобно воспользоваться известными результатами теории солитонов \cite{Kaup,KarpmanMaslov,Newell}. 
\par
Будем искать убывающее при $|z|\to\infty$ решение возмущенного уравнения (\ref{perturbedNShE}) с главным членом в солитонном виде:
$$
\psi_0=\frac{2i\eta\exp(-2i\kappa z-i\Omega)}{\cosh(2\eta z+V))}.
$$
Параметры солитона будем искать в виде:
\begin{eqnarray*}
 \eta(\sigma)\sim\eta_0+\mathcal{O}(\sigma^{-1/2}),\quad 
\kappa(\sigma)\sim\kappa_0+\mathcal{O}(\sigma^{-1/2}),\\
\Omega\sim 4\int_0^\sigma (\kappa^2+\eta^2)d s+\Omega_1(\sigma),\quad \Omega_1(\sigma)\sim\mathcal{O}(\sigma^{1/2}),\\
V\sim 8\int_0^{\sigma}\kappa\eta d s +V_1(\sigma),\quad V_1(\sigma)\sim\mathcal{O}(\sigma^{1/2}).
\end{eqnarray*}
\par
Для того, чтобы вычислить параметры возмущенного солитона с указанной здесь точностью, можно воспользоваться двумя законами сохранения. С помощью стандартной процедуры выведем уравнения для эволюции параметров. Рассмотрим сопряженные уравнения:
\begin{eqnarray}
\partial_\sigma\phi-i\partial_{zz}\phi-2i|\phi|^2\phi-ih=0,\label{forsedNShE}\\
\partial_\sigma\overline{\phi}+ i\partial_{zz}\overline{\phi}+2i|\phi|^2\overline{\phi}+i\overline{h}=0.\label{conjugForcedNShE} 
\end{eqnarray}
Умножим первое уравнение на $\overline{\psi}$, второе  на $\psi$, просуммируем полученные уравнения и проинтегрируем по всей вещественной оси $z$. Предполагая убывание решения при больших $|z|$ и принадлежность решения классу $L_2$, получим:
\begin{eqnarray*}
 \frac{d}{d s}\int_{-\infty}^{\infty}|\phi|^2d z-i\int_{-\infty}^{\infty}(h\overline{\phi}-\overline{h}\phi )d z=0.
\end{eqnarray*}
Поставим формулу для главного члена асимптотики в интегральное тождество? в результате:
\begin{equation}
 4\frac{d\eta}{d s}\sim i\int_{-\infty}^{\infty}(h\overline{\phi}-\overline{h}\phi )d z
\end{equation}
Из асимптотической формулы для $\eta$ и свойств ортогональности решений линеаризованного уравнения Шредингера \cite{Newell} следует, что:
\begin{equation}
H_0=i\int_{-\infty}^{\infty}(h\overline{\phi_0}-\overline{h}\phi_0 )d z\sim o(s^{-1}).
\label{asympEqForEvolutionEta}
\end{equation}
Обозначим через $h$ невязку:
$$
h= i\partial_{\sigma}\phi_0+\partial_{zz}\phi_0+2|\phi_0|^2\phi_0+\frac{\cal F}{\sigma^{3/4}}\exp(i\sigma)-i\frac{\phi_0}{4\sigma}.
$$
Вычислим интеграл (\ref{asympEqForEvolutionEta}):
\begin{eqnarray*}
H_0\sim \frac{\eta}{\sigma}+\frac{i}{\sigma^{3/4}}\int_{-\infty}^\infty {\cal F} e^{i\sigma}\frac{-2i\eta e^{2i\kappa z+i\Omega}}{\cosh(2\eta z+V)}-\overline{{\cal F}} e^{-i\sigma}\frac{2i\eta e^{-2i\kappa z-i\Omega}}{\cosh(2\eta z+V)}\sim
\\
\frac{\eta}{\sigma}-\frac{2|f|\eta}{\sigma^{3/4}}\cos\big(\sigma+\arg({\cal F})- \kappa V/\eta+\Omega)\int_{-\infty}^\infty\frac{\cos(\kappa X/\eta)}{\cosh(X)}\frac{dX}{2\eta}
\\
= \frac{\eta}{\sigma}-\frac{\pi|{\cal F}|\cos\big(\sigma+\arg({\cal F})- \kappa V/\eta+\Omega\big)}{\sigma^{3/4}\cosh\left(\frac{\pi \kappa}{2\eta}\right)}.
\end{eqnarray*}
Обозначим $A=|{\cal F}/\sigma^{1/4}|$. Требование $H_0\sim0$ приводит к двум условиям. Во-первых, отсутствию осцилляций:
$$
\alpha\equiv\arg({\cal F})+\sigma-\kappa V/\eta+\Omega\sim \mathcal{O}(1),
$$
что в главном дает:
$$
4(\kappa_0^2+\eta_0^2)=1.
$$
Во-вторых, асимптотическое равенство дает:
$$
\eta_0=-\frac{A\pi\sin(\alpha)}{\cosh\left(\frac{\pi \kappa_0}{2\eta_0}\right)}.
$$
Еще одно уравнение для вычисления параметров солитона получается из закона сохранения импульса для нелинейного уравнения Шредингера. Умножим (\ref{forsedNShE}) на $\partial_z\overline{\phi}$, уравнение (\ref{conjugForcedNShE}) на $\partial_z\phi$. Вычтем из первого полученного выражения второе и с помощью интегрирования по частям получим:
\begin{equation}
 \frac{d}{d \sigma}\int_{-\infty}^{\infty}\partial_z\phi\overline{\phi}d z+i\int_{-\infty}^{\infty}(h\partial_z\overline{\phi}+\overline{h}\partial_z\phi)d z.
\label{conservationLaw2}
\end{equation}
В формулу (\ref{conservationLaw2}) подставим выражение для главного члена асимптотики, в результате, так же как и выше, получим:
\begin{equation}
 8\frac{d}{d s}(\eta\kappa)=\int_{-\infty}^{\infty}(h\partial_z\overline{\phi}+\overline{h}\partial_z\phi)d z.
\label{EqForEvilutionEtaKappa}
\end{equation}
Подстановка асимптотических формул для $\eta,\kappa$ дает:
\begin{equation}
 H_1\sim\int_{-\infty}^{\infty}(h\partial_z\overline{\phi_0}+\overline{h}\partial_z\phi_0)d z\sim o(s^{-1}).
\label{asympEqForEvolutionKappa}
\end{equation}
\par
Подставим формулу для $h$ в (\ref{asympEqForEvolutionKappa}). Учитывая формулу для производной
$$
\partial_z\phi_0=-2i\kappa\phi_0-2\eta\phi_0\tanh(2\eta z+V)),
$$
получим:
\begin{eqnarray*}
H_1=\int_{-\infty}^\infty\left(-2i\kappa h\overline{\phi}+2i\kappa\overline(h)\phi\right)dz + \\
\int_{-\infty}^\infty\left(-2\eta h\overline{\phi}\tanh(2\eta z+V)-2\eta\overline{h}\phi\tanh(2\eta z+V)\right)dz=\\
-2\kappa H_0+\int_{-\infty}^\infty\left(-2\eta h\overline{\phi}\tanh(2\eta z+V)-2\eta\overline{h}\phi\tanh(2\eta(z+4\kappa s+V_1))\right)dz.
\end{eqnarray*}
Из $H_0\sim0$ следует:
$$
H_1\sim\frac{1}{\sigma}\int_{-\infty}^{\infty}Ae^{i\sigma}\left(\frac{4i\eta^2e^{2i\kappa z+\Omega}}{\cosh(2\eta z+V)}\tanh(2\eta z+V)\right)+
$$
$$
Ae^{-i\sigma}\left(-\frac{4i\eta^2e^{-2i\kappa z-\Omega}}{\cosh(2\eta z+V)}\tanh(2\eta z+V)\right)=
$$
$$
4\eta A\sin(\alpha)\int_{-\infty}^\infty \frac{\sin(\kappa X/\eta)\tanh(X)}{\cosh(X)}dX.
$$
Тогда:
$$
H_1\sim-4\kappa \eta\frac{A\pi\sin(\alpha)}{\cosh\left(\frac{\pi\kappa}{2\eta}\right)}.
$$
Или
$$
\frac{\kappa_0 A\sin(\alpha)}{\cosh\left(\frac{\pi \kappa_0}{2\eta_0}\right)}\sim0.
$$
В результате получим систему уравнений для параметров солитона:
\begin{eqnarray*}
\alpha\sim\{0,\pi\},\\
\eta_0=\pm\frac{A}{\cosh\left(\frac{\pi \kappa_0}{2\eta_0}\right)},\\
4(\kappa_0^2+\eta_0^2)=1.
\end{eqnarray*}
Обозначим: $\mu=\kappa_0/\eta_0$ в результате получим:
\begin{equation}
\eta_0=\pm\frac{1}{\sqrt{2(\mu^2+1)}},\quad A=\frac{\cosh\left(\frac{\pi \mu}{2}\right)}{\sqrt{2(\mu^2+1)}}
\end{equation}

\section{Влияние диссипации}
\par
Рассмотрим уравнение (\ref{primResEq}) при $\nu>0$. Для функции $\phi$  и переменных $\sigma,z$ исходное уравнение примет вид:
\begin{equation}
 i\partial_\sigma\phi+\partial_{zz}\phi+|\phi|^2\phi +\frac{\cal F}{\sigma^{3/4}}e^{i\sigma}+i\frac{\nu}{2\sqrt{\sigma}}\phi+i\frac{\phi}{4\sigma}=0.
\end{equation}
В этом случае диссипативное слагаемое с параметром $\nu$ и внешняя сила ${\cal F}$ дают уравнения для параметров солитона в виде:
$$
\frac{2\nu\eta_0}{\sqrt{\sigma}}+\frac{1}{\sigma^{3/4}}\frac{\pi|{\cal F}|\cos(\alpha)}{\cosh\left(\frac{\pi\kappa_0}{2\eta_0}\right)}\sim0.
$$
Откуда следует, что при $\sigma\to\infty$ и постоянной амплитуде $\cal F$ параметр $\eta_0$ убывает как $\sigma^{-1/4}$. Следовательно, построенная асимптотика $\Psi=\mathcal{O}(1)$ при $\tau\to\infty$. 

Рост асимптотического решения сохраняется, если растет амплитуда вынуждающей силы, а именно при ${\cal F}=\mathcal{O}(\nu\sigma^{1/4})$ или как $\mathcal{O}(\sqrt{\tau})$ в терминах исходной переменной.
\par
Необходимость роста амплитуды возбуждения для сохранения авторезонанса для решений другого типа была известна ранее \cite{SKKS,KS}.

\section{Заключение}
\par
Для уравнения главного резонанса в частных производных (\ref{primResEq}) (неавтономного нелинейного уравнения Шредингера c внешней силой) построен главный член растущего, как $\sqrt{\sigma}$, асимптотического решения в солитонной форме. Оказалось, что для авторезонансного роста солитона достаточно внешней силы с убывающей, как $\mathcal{O}(1/\sqrt{\tau})$, амплитудой. Показано, что при диссипации для растущего асимптотического решения необходим рост внешней силы как $\mathcal{O}(\nu \sqrt{\tau})$.

{\bf Благодарности}
Выражаю признательность С.Г. Глебову за многочисленные обсуждения. Работа выполнена при поддержке РФФИ 11-01-91330.

\end{document}